# Performance Analysis of Two-Hop Cooperative MIMO transmission with relay Selection in Rayleigh Fading Channel

Ahasanun Nessa[1], Qinghai Yang[2], Sana Ullah[1], Md Huamaun Kabir[1], Kyung Sup Kwak[1]

[1]Graduate School of Information Technology and Telecommunications, Inha University,
#253 Yonghyun-dong, Nam-gu, Incheon, 402-751, Korea. Email: anessa_2006@yahoo.com

[2]School of Telecommunications Engineering, Xidian University, No.2 Taibainan-lu, Xi'an, 710071,
Shaanxi, China. Email: qhyang@xidian.edu.cn

**Abstract**— Wireless Relaying is one of the promising solutions to overcome the channel impairments and provide high data rate coverage that appears for beyond 3G mobile communications. In this paper we present an end to end BER performance analysis of dual hop wireless communication systems equipped with multiple Decode and Forward relays over the Rayleigh fading channel with relay selection. We select the best relay based on end to end channel conditions. We apply Orthogonal Space Time Block coding (OSTBC) at source, and also present how the multiple antennas at the source terminal affects the end to end BER performance. This intermediate relay technique will cover long distance where destination is out of reach from source.

**Keywords**-Bit error rate (BER), amplify and forward (AF), multiple input multiple output (MIMO), decode-and-forward (DF), probability density function (PDF).

## I. INTRODUCTION

Dual hop transmission is a technique by which the channel from source to destination is split into two possibly shorter links using a relay. In this case the key idea is that the source relays a signal to destination via a third terminal that acts as a relay [1]. It is an attractive technique when the direct link between the base station and the original mobile terminal is in deep fade or heavy shadowing or when the destination is out of reach of the source.

Cooperative relaying is a promising extension to relay networks where several relay stations transmit jointly to same destination yielding diversity gain [2]. Depending on the nature and complexity of the relays cooperative transmission system can be classified into two main categories, namely regenerative or non regenerative systems. The performance of both systems has been well studied in [3], [4], [5]. In [5] it is shown that outage probability may be reduced via a variety of cooperation protocols. In [5] and [6] various protocols for ergodic capacity have been analyzed but they all are for fixed single relay system with AF (Amplify-and-Forward) and DF (Decode-and-Forward) modes. But all this analysis all nodes have only one antenna consideration. Dual hop transmission of multiple antenna equipped relays has been analyzed in paper [7]. In this paper, our aim is to analyze the system with multiple relay nodes where each relay node has one transmit and one receive antenna and source transmitting data using more than one transmit antenna . We assume the two hop experience independently, not necessarily identically distribute Rayleigh fading reliability by averaging over independent channel realization and would be more efficient to combat fading and shadowing.. Moreover in this paper at the receiver end we are not considering all received signal passing through different relay stations. We are selecting signal from one of the best relay stations to assist the communication like a single source destination pair. It will reduce the decoding complexity at receiver side and also in the same time will achieve diversity gain.

The paper is outlined as follows: Section II introduces the system and channel model. Section III derives the probability density function and moment generation function of the received SNR per bit and analyzes the BER performance when M-ary PSK constellations are used. Relay selection protocol is described in Section IV. Simulation results of our end to end BER performance are presented in Section V. Finally Section VI presents conclusion and future work.

## II. SYSTEM AND CHANNEL MODEL:

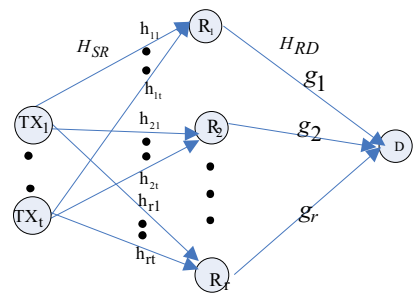

Figure 1. Two-hop relay network with antenna arrays at source and r relays are relaying signal to D.

Consider a wireless network with $r$ relay nodes which are placed randomly and independently according to some distribution. The source is equipped with multiple transmit antennas and each relay node has a single antenna which can be used for both transmission and reception. After receiving the signal from source each relay node decodes it and among them only one transmits it to destination. We assume there is no direct link from source to destination. We are applying OSTBC at the source. No channel information is available at source. So no power or bit loading is performed at source. Each transmit antenna of source is assumed to use the same transmit power $\sigma_s^2 = \frac{P}{t}$, where $P$ is the total transmission power of the source and $t$ is the number of antennas at source. At first time slot the source transmits the signals over the uplink matrix channel $H_{SR}$ to the relays. After receiving the signal each relay decode the signal and find out the best relay among them and then transmits it to destination.

## III. BER ANALYSIS

We are considering the dual hop wireless communication in which source equipped with multiple antennas and communication with destination via number of relay nodes. In order to achieve spatial diversity we are applying OSTBC at source. We assume there are $r$ relays and number of transmit antennas at source is t. So the channel matrix for the first hop is given by

$$H^{SR} = \begin{pmatrix} h_{11} & .. & h_{1t} \\ \vdots & h_{ij} & \vdots \\ h_{r1} & .. & h_{rt} \end{pmatrix} \quad (1)$$

where the element $h_{ij}$ denotes the channel gain between the $i$ th relay and the $j$ th transmit antenna of source, $i=1,2,...r$ and $j=1,2,...t$. We assume that each element of $H^{SR}$ is an independent and identically distributed complex Gaussian random variable with zero mean and $\beta_1$ variance. If we notice carefully we observe each row of $H^{SR}$ represents the channel coefficient between source and relay. So the channel matrix for each relay can be represented by

$$\alpha_i = (h_{i1} \quad . \quad h_{it}) \quad \text{for i=} i=1,2,...r$$

And for the second hop $g_i$ is the individual relay to destination fading amplitude.

When OSTBC X is used at the source the signals received at each relay are given by

$$Y^i = \alpha^i X + E^i \quad (2)$$

where

$$Y^i = \begin{bmatrix} y_1^i . & .y_L^i \end{bmatrix} \quad (3)$$

and

$$E^i = \begin{bmatrix} e_1^i . & .e_L^i \end{bmatrix} \quad (4)$$

$y_l^i$ and $e_l^i$ denote the received signal and the additive complex white Gaussian noise with mean zero and variance $\sigma_A^2$ respectively, at the $i$th relay during the $l$ th symbol duration where the block length of the OSTBC is denoted by $L$. After receiving the signal each relay decodes them with an efficient ML detector and take decision among them who is more opportunistic for relaying signal to destination and then transmit it to destination. We are discussing the relay selection protocol in Section IV. Let $i$ th relay is the most appropriate the for relaying signal to destination .Then the received signal at destination is

$$Y^D = g_i X_1 + e^D \quad (5)$$

where $X_1$ is the transmitted signal matrix from relay to destination and $e^D$ is the complex additive white Gaussian noise with mean zero and variance $\sigma_B^2$. Let $\sigma_A^2 = \sigma_B^2 = \sigma^2$. From (2) the received SNRs at each relay for the first hop can be given as

$$\gamma^R(\rho) = c\rho |\alpha^i|^2 = = c\rho \sum_{j=1}^{t} |h_{ij}|^2 \quad (6)$$

where $c = L/(t.K.Log_2 M)$ and $\rho = P/\sigma^2$. K is number of complex signal is transmitted by OSTBC code X. And for the second hop it is like a SISO channel. For M order modulation the SNR of second hop can be given by

$$\gamma^D(\rho) = \rho |g_i|^2 / \log_2 M \quad (7)$$

So the MGFs of the first hop and second hop for $\gamma^R(\rho)$ and $\gamma^D$ are respectively obtained by

$$M_{\gamma^R}(s) = (1 + c\beta_1 \rho s)^{-t} \quad (8)$$

$$M_{\gamma^D}(s) = 1/(1 + \gamma^D s) \quad (9)$$

By taking the inverse Laplace transform of (8) and (9) The PDFs of $\gamma^R(\rho)$ and $\gamma^D(\rho)$ are respectively given by

$$f_{\gamma^R(\rho)}(\gamma) = (c\beta_1\rho)t \frac{1}{(t-1)!} e^{-\gamma/c\beta_1\rho} \quad (10)$$

$$f_{\gamma^D(\rho)}(\gamma) = \frac{1}{\gamma} e^{-\gamma/\gamma^D(\rho)s} \quad (11)$$

**BER of M-ary PSK constellation** : The PDF of phase $\theta$ of the received signal with SNR $\gamma$ is given by in [10]

$$f_\theta(\theta|\gamma) = \frac{1}{2\pi} e^{-\gamma \log_2 M} \left[ 1 + \cos\theta \sqrt{4\pi \log_2 M \gamma} e^{\gamma \log_2 M \cos^2\theta} \left(1 - \frac{1}{2} erfc\left(\sqrt{\gamma \log_2 M} \cos\theta\right)\right) \right]$$
(12)

where $\text{erfc}(x) = \frac{1}{\sqrt{\pi}} \int_x^\infty e^{-y^2} dy$. Then the exact probability that the phase $\theta$ of the received signals lies in a decision region $[\theta_l, \vartheta_u]$ is given by

$$\Pr\{\theta \in [\theta_l, \vartheta_u]; \rho\} = \int_0^\infty \int_{\theta_l}^{\vartheta_u} f_\theta(\theta|\gamma) f_{\gamma(\rho)}(\gamma) d\theta d\gamma \qquad (13)$$

The BERs of M-ary PSK constellation for the first and second are respectively obtained by

$$P_R^{M-PSK}(\rho) = \frac{1}{\log_2 M} \sum_{j=1}^M e_j \int_0^\infty \int_{\theta_{l_j}}^{\theta_{u_j}} f_\vartheta(\theta|\gamma) f_{\gamma^R(\rho)}(\gamma) d\theta d\gamma \qquad (14)$$

$$P_D^{M-PSK}(\rho) = \frac{1}{\log_2 M} \sum_{j=1}^M e_j \int_0^\infty \int_{\theta_{l_j}}^{\theta_{u_j}} f_\vartheta(\theta|\gamma) f_{\gamma^D(\rho)}(\gamma) d\theta d\gamma \qquad (15)$$

where $\theta_{l_j} = (2j-3)\pi/M$ and $\theta_{u_j} = (2j-3)\pi/M$ for j=1,…M and $e_j$ is the number of bit errors in the decision region.

## IV. RELAY SELECTION

In this paper, all relay stations are not relaying signal to destination. At the receiver end we are only getting signal which are coming form the best station. We are assuming source to relay and relay to destination channel state information is available to each relay. The relay nodes monitor the instantaneous channel conditions. After receiving the signals, each relay decodes hem with an efficient ML detector and find out among them which one is more opportunistic for relaying signal to destination. The term opportunistic has been widely used in various different context. In previous work opportunistic relay is defined considering distance toward source or destination [9] or sometimes considering the channel condition [8]. The relay selection based on distance is not a good selection since communication link between transmitter and receiver locating in the same distance might have enormous difference in terms of received signal due to fading and shadowing. In this paper we are assuming all relays can listen to each other. After monitoring the instantaneous channel condition each relay also broadcast the information to other relay nodes. And they come to know among them which one is more opportunistic in this time.

In our study we denote the relay to destination channel state information $H_{SR}$ and $H_{RD}$. Let $\alpha_{si}$ and $\alpha_{id}$ is the channel state information to $i$th relay from source to $i$th relay and $i$th relay to destination respectively. The channel estimates $\alpha_{si}$, $\alpha_{id}$ describe the quality of the wireless path between source-relay -destination to each relay. Where $\alpha_{si}$ is calculated by relay $i$ by the following equation.

$$|\alpha_{si}| = (|h_{i1}| + . . . + |h_{it}|)/t . \qquad (16)$$

$|\alpha_{id}|$ is the fading amplitude from relay to destination. Since the two hops are both important for end to end performance each relay calculates corresponding $h_i$ based on the two decision rules and broadcast the measured values among them

$$\text{Rule 1: } h_i = \min\{|\alpha_{si}|^2, |\alpha_{id}|^2\} \qquad (19)$$

$$\text{Rule 2: } h_i = \frac{2}{\frac{1}{|\alpha_{si}|^2} + \frac{1}{|\alpha_{id}|^2}} = \frac{2|\alpha_{si}|^2|\alpha_{id}|^2}{|\alpha_{si}|^2 + |\alpha_{id}|^2} \qquad (20)$$

The relay i that maximizes function $h_i$ is one with the "best" end to end path between initial source to destination.

After discovered best relay then it relaying to destination. In this paper it is assumed the destination have perfect channel information available for decoding the received signal.

## V. SIMULATION RESULT

In this section, we are presenting our simulation result about BER performance. We consider QPSK and 16QAM constellation for 2 and 4 transmit antenna equipped source. We are assuming the channels are Rayleigh fading channel. Two sorts of simulation are performed one for decision rule1 and another for decision rule2. And performances are nearly same for both two cases.

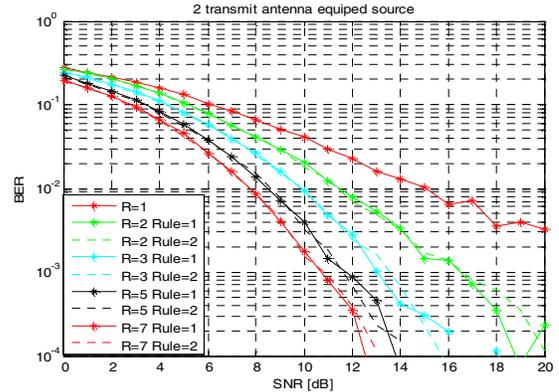

Figure 2. BER vs SNR for Relay selection. Tx=2,modulation=QPSK.

In Fig2 and Fig3 we consider 2 transmit antenna and 4 transmit antenna at source respectively. By comparing Fig 2 and Fig 3 we illustrate how BER performance can be improved by increasing the number of antennas and relays. If we select more relays than it is possible to improve BER performance instead of using more transmit antennas at source. It will reduce the cost of implementing more antennas at base station. Modulation order also affects the difference between the BER performances, that is shown in Fig 4 and Fig 5. If the modulation order gets higher, the difference becomes egligible. Comparing Fig 2 and Fig 4 and Fig 3 and Fig 5, it is easily noticeable.

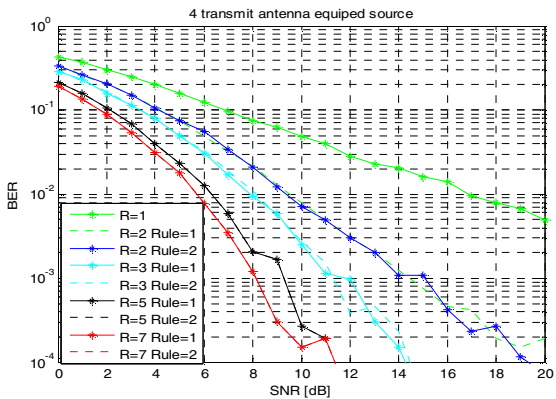

Figure 3. BER vs SNR for Relay selection. Tx=4,modulation=QPS

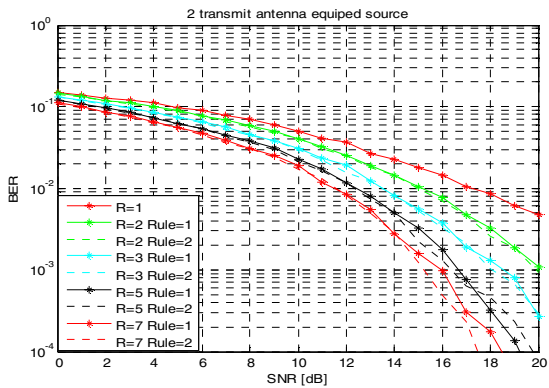

Figure 4. BER vs SNR for Relay selection. Tx=2,modulation=16QAM

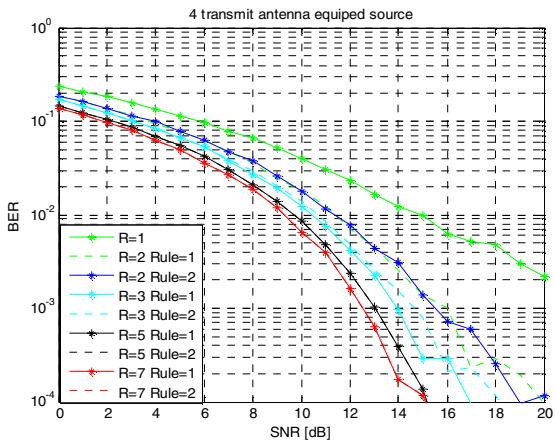

Figure5. BER vs SNR for Relay selection. Tx=4,modulation=16QAM

## VI. CONCLUSION

In this paper, we presented end to end BER performance of dual hop wireless transmission employing transmit diversity with OSTBC and receiver diversity by using multiple relays where the destination is out of reach from source. We show the BER performance by varying the number of antennas and number of relays. But this performance improvement is not too much if the channel is noisy. In our design we are only considering the best signals those are coming from best relay station for reducing receiver complexity. In future our work will be evaluated under more advanced receiver structure for realistic channel consideration and we will continue our work for multi hop transmission for covering long distance.

## ACKNOWLEDGEMENT

This research was supported by the MKE(Ministry of Knowledge Economy), Korea ,under the ITRC( Information Technology Research Center ) support program supervised by the IITA( Institute of Information Technology Assessment )" (IITA-2008-C1090-0801-0019)